# Structural signatures of strings and propensity for mobility in a simulated supercooled liquid above the glass transition

**Magnus N. J. Bergroth[1] and Sharon C. Glotzer[1,2] ***

*[1]Departments of Chemical Engineering and [2]Materials Science & Engineering*

*University of Michigan, Ann Arbor, Michigan 48109-2136*

*To whom correspondence should be addressed, sglotzer@umich.edu

**Submitted October 14, 2006**

**Abstract**

By molecular dynamics (MD) simulation of the one-component Dzugutov liquid in a metastable equilibrium supercooled state approaching the glass transition, we investigate the structural properties of highly mobile particles moving in strings at low temperature $T$ where string-like particle motion (SLM) is well developed, and compare the properties of the string particles to the properties of less mobile particles in the liquid. We find that SLM occurs most frequently in the boundary regions between clusters of icosahedrally-ordered particles and disordered, liquid-like, domains. Further, we find that the onset $T$ for significant SLM coincides with the $T$ at which clusters of icosahedrally-ordered particles begin to appear in considerable amounts, which in turn coincides with the onset $T$ for non-Arrhenius dynamics. Through analysis of the local potential energy, coordination numbers, the radial distribution function and common neighbor analysis for string versus bulk particles, we find a unique structural environment for strings that is different from the structure of the bulk liquid at any $T$. This unique string environment persists from the melting $T$ upon cooling to the lowest $T$ studied in the vicinity of the mode-coupling temperature, and is explained by the existence of rigid elongated cages. Over this *T*-range, the string particles exhibit significantly less icosahedral short-ranged order (SRO) relative to that of the bulk particles. We also form a criterion based solely on structural features of the local environment that allow the identification of particles with an increased propensity for mobility. We conclude from the similarities in properties found for particles satisfying the criterion in three different model liquids that that such a criterion can be universally implemented to identify particles with a propensity for high mobility, and we establish such a connection for the Dzugutov and a polymer system.

**I. Introduction**

The glass transition remains an open problem in condensed matter physics.[1-4] Despite vigorous investigation via experiments, theory, and computer simulations, a complete understanding of the microscopic details governing the glass transition is still lacking. However, over the last several years much progress has been made in understanding the nature of glassy and supercooled liquids, both with respect to structure and dynamics.

In an early communication regarding the structure of materials, Frank[5] pointed out that the energy of an icosahedral cluster of 13 Lennard-Jones atoms is about 8.4% lower per particle than that of periodic (e.g. fcc or hcp) clusters with 13 atoms. Based on this finding he speculated that icosahedral short-range order (SRO) made possible the experimentally observed supercooling of liquid metals. Indeed, Frank's speculation about the importance of icosahedral SRO for supercooling was later confirmed in computer simulations of supercooled Lennard-Jones model systems.[6-8] Recent simulation studies also show that icosahedral SRO in metallic liquids increases with the degree of supercooling, see for example Refs. 9 and 10. Experimental evidence also exists that show icosahedral SRO to be present in supercooled liquids; for example, by scattering of totally internally reflected X-rays, Reichert[11] observed five-fold local symmetry in liquid lead adjacent to a silicon wall; using a combination of electromagnetic levitation and neutron scattering, Schenk[12] and coworkers found icosahedral SRO to occur in the bulk metallic melt independent of the system investigated, and that the SRO strongly increased with the degree of supercooling. In addition, in a more recent study[13] on Al- Fe-Co liquids it was established that topological icosahedral SRO is present in those liquids. Holland-Moritz[14] has also found direct experimental proof of icosahedral short-range order in deeply supercooled bulk liquids of a large variety of pure metals and metallic alloys.

A fairly recent discovery that contributes to our understanding of the dynamics of supercooled liquids is the notion of spatially heterogeneous dynamics (SHD) (see, e.g. Refs. 15-22) where subsets of neighboring particles are found to be significantly more or less mobile over a given time interval, compared to the average particle, than would be expected from a Gaussian distribution of particle displacements[19]. A Gaussian distribution is expected when relaxation processes are inertial or diffusive, as they are at very short and very long times in supercooled liquids. The cage-breaking process, which occurs at intermediate times, involves the dynamically correlated motion of neighboring particles[25]. Using fluorescent confocal scanning laser microscopy, Weeks and coworkers[23] as well as Kegel and van Blaaderen[24] made direct observation of SHD in colloidal hard sphere liquids. It has further been found that correlated particle motion in model glass-formers occur in 1D string-like paths. This has been observed for various model systems such as Lennard-Jones,[25,26] Dzugutov,[27] colloids,[28] water,[29,30] and polymers.[31,32] In a BKS-silica study,[33] SLM was also detected but to a far lesser extent and only for oxygen. Experimentally, SLM has been observed in colloidal liquids, in both quasi-2D[34,35] and in 3D[36] systems.

Recently Widmer-Cooper and Harrowell[37] made significant progress in establishing a connection between structure and dynamics. They used a particle based 2-D model to develop a technique that makes possible identification of regions with a high propensity for mobility. In short, their idea is to restart several MD runs from the same initial configuration, while assigning different initial momenta, thereby creating different subsequent system trajectories. In doing so, they were able to establish a link between a given configuration and subsequent dynamics, thereby introducing the concept of propensity for mobility. Jain and de Pablo[38] used a Lennard-Jones system to show a connection between structure and dynamics in that a higher degree of local distortion, measured in terms of the tetrahedricity of the Delaunay simplices,[39] was noted for particles that were likely to transition faster to a neighboring metabasin on the potential energy landscape. This result corresponds to that of Starr et al.,[40] who found for a polymer melt a broad range of values in the distribution of asphericity of the Voronoi[39] volumes. Recent work by Matharoo et al.[41] on simulations of ST2 water shows a clear correlation between particles with propensity for low energy (a structural feature) and those with propensity for low mobility (a dynamical feature).

We present in this manuscript results from a MD simulation of a monatomic liquid modeled by the Dzugutov pair potential,[42] with the aim of identifying and correlating structural and dynamical features that will aid in gaining further understanding of the propensity for high mobility and the origin of SLM as liquids approach the glass transition. We carry out detailed analyses of the local environment of particles moving in strings, and contrast and compare this environment to that of particles in the bulk liquid, in particular particles involved in icosahedral ordering. We characterize the local particle environment by determining various quantities, such as potential energy distributions, distributions of distances to nearest neighbors, and coordination numbers. We also demonstrate that SLM frequently occurs in the neighborhood of clusters of icosahedrally-ordered particles. We divide the particles into different "particle types", where the particle types indicate if a particle is located in a region that is icosahedrally-ordered, more liquid-like (i.e. disordered), or in the boundary between these regions. A precise definition of these types is given in the results section. In addition, we quantify the local structure in more detail using common neighbor (CN)[6,43-45] analysis, and relate these results to the occurrence of SLM. We find that a certain local structural environment, characterized by low CN numbers, favors SLM, and we form a criterion based solely on structural features that signals increased propensity[37] for particle mobility.

This manuscript is organized as follows. In Sec. II we introduce the model employed for the main part of this study. We also provide a brief summary of some characteristic features of the model and refer the reader to other works that use the model. In Sec. III we revisit the notion of SLM and the local structure of the liquid. Sec. IV analyzes the local environment of the liquid based on grouping particles into different "types" based on the underlying icosahedra structure. In Sec. V we compare and contrast local structural features for particles in the bulk liquid to particles moving in strings. We then expand the structural study in Sec. VI by employing the CN analysis method.[6,43-45] In Sec. VII we form a criterion based solely on local structure that identifies particles with a propensity[37] for mobility. In Sec. VIII we analyze properties of the particles that satisfy the criterion in three different model liquids. Finally, we conclude in Sec. IX with a summary of our results.

**II. Model and simulation**

In this study we use the one-component Dzugutov (DZ) interaction pair potential, which is designed to favor icosahedral ordering in the first neighbor shell, and which was originally developed as a model of simple glass-forming liquid metals.[42] This model liquid has been successfully used in various works ranging from, e.g., studies of clustering structures and quasicrystals,[46-48] and spatially heterogeneous dynamics[49] to dynamic facilitation[50] and string-like motion.[27]

The Dzugutov potential is constructed to suppress crystallization common to most monatomic systems by the introduction of a repulsive term representing the Coulomb interactions that are present in a liquid metal. This term gives rise to a second maximum that is strategically located to hinder particles residing in the second neighbor shell from finding energetically favorable sites as in a face-centered (fcc) or body-centered (bcc) cubic configuration.[51] By introducing frustration locally in this way, crystallization into normally preferred crystal structures is disfavored, and the system remains in the supercooled liquid state longer before eventually crystallizing into a dodecagonal quasicrystal (a metastable state) or a stable crystalline (fcc) solid phase. In all, one attractive and two repulsive regimes characterize this potential, as shown in Figure 1, where the Dzugutov potential is plotted together with the Lennard-Jones (LJ) potential for comparison.

The explicit form and the parameters of the Dzugutov potential are given below.

$$U_{Dz} = U_1 + U_2$$
$$U_1 = A(r^{-m} - B)\exp\left(\frac{c}{r-a}\right), \quad r < a$$
$$U_1 = 0, \quad r \geq a \quad (1)$$
$$U_2 = B\exp\left(\frac{d}{r-b}\right), \quad r < b$$
$$U_2 = 0, \quad r \geq b,$$

where A = 5.82, B = 1.28, a = 1.87, b = 1.94, c = 1.1, d = 0.27, and m = 16.

Dimensionless units are used in the simulation, where the energy has units of $\varepsilon$, the temperature $T$ has units of $\varepsilon/k_B$, and $k_B$ is the Boltzmann constant. Length has units of $\sigma$, and the dimensionless time $t$ has units of $\sigma(m/\varepsilon)^{1/2}$.[42] The mass $m$ and energy $\varepsilon$ are set to unity. Our system contains $N$ = 17576 particles confined with periodic boundary conditions in a cubic box with a fixed density of 0.85. The simulations are performed in the NVT (canonical) ensemble, and the temperature is maintained using a Berendsen thermostat.[52] The integration scheme used is velocity Verlet,[53] with a time step of 0.01$t$. The onset $T$ where this DZ model enters a supercooled, non-Arrhenius regime on cooling is at $T_o \approx 0.8$[48] and the critical $T$ of the mode-coupling theory[54] is $T_{MCT} \approx 0.4$.[49] Melting upon heating occurs at $T_M \approx 0.7$,[55] and crystallization at $T_{cryst} \approx 0.5$.[48] We start our simulation at $T$ = 1.6, where we equilibrate the system for a sufficient time to remove any "memory" of the initial configuration. We then cool the system in a stepwise manner and equilibrate at each $T$ before collecting data. We have through careful monitoring of our simulations ensured that the system is quiescent on the timescales of our study. The results presented below are obtained in the usual way by time averaging over a trajectory and, in addition, by averaging over several independent simulations for improved statistics. Please refer to Ref. 27 for a detailed description of the SHD characteristics of this model system.

**III. Strings and icosahedral structure**

Strings formed by particles replacing one another in quasi one-dimensional paths are defined here identically to Ref. 25, that is, by connecting two mobile particles $i$ and $j$ if $\min[|\vec{r}_i(t) - \vec{r}_j(0)|, |\vec{r}_i(0) - \vec{r}_j(t)|] < 0.6$. Mobile particles are defined as in Ref. 19, i.e. by selecting the top 5% of the particles with the largest displacement over a given time interval $t^*$. Following Ref. 27, the time $t^*$ is chosen such that the string length is maximized at a given $T$.

In Figure 2 we show a plot of selected string length distributions for a broad range of $T$'s. $T = 0.42$ is the coldest $T$ attained in this study and here string-like motion is most clearly pronounced. At higher $T$ the strings become shorter and fewer in number.

Included in Figure 2 are distributions of strings consisting of three or more particles. On closer inspection we find that, as the strings grow in size, the distribution for $T = 1.0$ drops off more rapidly than the distributions for lower $T$. We find that string-like motion is more pronounced, and that longer strings occur much more frequently in the system for $T = 0.8$ and below, which coincides with the onset $T$ for non-Arrhenius dynamics ($T_o \approx 0.8$). Donati et al.[25] speculated that the lengths of the strings are important for the overall dynamics of the system. In previous work[27], it was found that SLM constitutes an important feature in the relaxation of the mobile domains in the DZ model liquid. It was also discovered that the string length ($l$) is important in determining the type of motion the string undergoes. Only short strings ($l < 7$) exhibited a coherent type of motion. Larger strings were found to be comprised of sub-units, labeled as micro strings, within which the motion is coherent. Here coherent motion is referred to as when a particle is replacing another within a small time interval. In addition, we have shown in a previous paper[25,56] that the strings are good candidates for the cooperatively rearranging regions of Adam and Gibbs,[57,58] which indicates that a thorough understanding of the nature of the strings is important for obtaining a complete microscopic picture of the origin of SHD and the increasingly slow relaxation of liquids approaching their glass transition.[25]

To relate string-like motion to local structure, we have developed a simple but effective analysis method (to be defined shortly) that allows us to determine exactly where particles moving in strings are located relative to the clusters of icosahedrally-ordered particles known to be present[42,48] in this model liquid. This method is based on the geometry of an icosahedron. Following Dzugutov,[42] we define icosahedra to be clustered if three or more particles are shared between two neighboring icosahedra. An example of a cluster of icosahedrally-ordered particles at $T = 0.42$ is shown in Figure 3.

We refer readers interested in these clusters to a manuscript in preparation that will present a detailed study of polytetrahedral clusters in the DZ liquid.[59] For this study we only note that our results are consistent with previous findings for the DZ model liquid;[48] we find clusters of icosahedrally-ordered particles to grow in size with decreasing *T*, and the total number of particles present in these clusters also to increase with decreasing *T*; this last observation is shown below in Figure 4 (top panel) and discussed in the context of particle "types", which we define next.

**IV. Characterizing the local structure using particle types**

We define here a method developed to analyze the local environment of the liquid based on grouping particles into different "types". The particle types used to analyze the local structure are determined by first finding all particles that reside at the center of an icosahedron. Next we group interconnected icosahedra together into clusters of icosahedrally-ordered particles as described above. We then label particles at the centers of icosahedra that are not part of a larger cluster as type *A* particles. The 12 nearest neighbors (*NN*) to the *A* particles are labeled *B*. Center particles that reside in icosahedra that are part of a larger cluster of icosahedrally-ordered particles are labeled *C*, and the 12 *NN* particles to the *C* particles are labeled *D*. We define one more particle type for icosahedrally-ordered particles that does not fit into any of the previous types as *E*. An *E* particle might be buried within a large cluster of icosahedrally-ordered particles, or it might be "trapped" between different such clusters. Occurrence of particle types defined so far indicates icosahedral SRO to be present in the system. The remaining particles are more "liquid-like" in nature, as they reside in non-ordered regions that are not part of the icosahedral structure. These particles are divided into two different types as follows: if a liquid-like particle is bordering icosahedrally-ordered particles, it is labeled as an *F* particle, and if not, we assign it to be a *G* particle. The definitions of the particle types are summarized below:

*A* – Center particles of icosahedra that are *not* part of a larger cluster

*B* – Nearest neighbors to *A*

*C* – Center particles of icosahedra that are part of a larger cluster

*D* – Nearest neighbors to *C*, and located on the surface of a larger cluster

*E* – Particles within larger clusters of icosahedra, but not *A*, *B*, *C*, or *D*

*F* – Liquid-like particles bordering an icosahedrally-ordered region

*G* – Liquid-like particles not bordering an icosahedrally-ordered region

Based on these particle types we obtain the frequency of occurrence plot shown in Figure 4.

From the top panel in Figure 4 we find that the local structure in the bulk liquid varies significantly with temperature. At high *T* the liquid-like *G* particles dominate the profile, indicating that the majority of the bulk liquid is liquid-like and that the degree of icosahedral SRO is low. As *T* is lowered, the *G* particles decrease in numbers and the degree of icosahedral SRO increases, with a rapid increase for $T < 0.5$. This is also reflected by the fast increase in the number of *D* particles and the decrease of *F* particles at $T < 0.5$. The shape of the *F* curve also reflects the growing number of clusters of icosahedrally-ordered particles in the bulk liquid with decreasing *T*; initially, as *T* decreases from high *T*, the number of *F* particles increases as the clusters start to grow in number and size and the total surface area of the clustered icosahedra therefore increases. Then, at $T \approx 0.5$ the number of *F* particles starts to decrease, as the clusters now grow in size and the total surface area of all these clusters start to decrease as small clusters merge into larger ones. Further, the presence of *B* at high $T$ indicates that icosahedra are always present in the bulk liquid, as these particles reside in non-connected icosahedra.

Also from the top panel in Figure 4, we find that it is at the onset of non-Arrhenius dynamics, $T_o \approx 0.8$, that clusters of icosahedrally-ordered particles first start to appear in a significant amount. This is shown by the appearance (upon cooling) of $D$ particles at $T = 0.8$; this is a reliable indicator of the appearance of clusters as the $D$ type particles reside on the surfaces of clustered icosahedra. This finding agrees well with an earlier study of icosahedral clusters in the inherent structures of the DZ system,[48] where an increase of the maximum cluster size was found to begin in the vicinity of $T = 0.8$.

From the lower panel in Figure 4, we immediately see that the particle type profile for the strings is very different compared to that of the bulk. For particles moving in strings, the liquid-like $F$ particles that border the icosahedrally-ordered clusters clearly dominate the profile at all $T$ covered in the string study, similarly to the bulk in this $T$-range, but for strings they continuously increase in occurrence with decreasing $T$. This result shows that strings are most commonly located close to the surfaces of clusters of icosahedrally-ordered particles, and that strings become more restricted to this location as $T$ is lowered. We show in Figure 5 an example of a typical string (in red) that extends through a liquid-like domain (light grey particles) located between clusters of icosahedrally-ordered particles (blue).

The string shown in Figure 5 constitutes a typical example of a string found in the DZ liquid. In this case, the string consists of 10 particles (in red). Note how 9 of the string particles are located in the immediate proximity of clusters of icosahedrally-ordered particles (shown in blue).

Continuing our discussion of Figure 4, note that the *D* particles move in strings with increased frequency as *T* is lowered. This indicates that, as the clusters grow in size, the surface particles become less tied to the clusters and can more easily become mobile; a phenomena that we deduce would arise from increasing frustration on the cluster surfaces as the clusters grow in size.[60,61] As Hoare[62] proposed, the size of non-crystallographic clusters, such as those formed from connected/interpenetrating icosahedra, will grow in size with decreasing temperature, but size limitations are expected to arise due to frustration as five-fold symmetry cannot span 3D space. We find our results to resemble those of Cui[35] who finds that the more mobile particles in the disordered boundary between ordered and disordered regions in a 2D colloidal system generate "string-like channels."

**V. Structural quantities: comparing between the bulk liquid and strings**

We next calculate the potential energy (PE) for particles in the bulk liquid and particles in strings and compare these measures. Calculations are performed at *T* = 0.42, 0.43, 0.46, 0.49, and 0.52. Over this *T*-range, which we will be referring to as the "low *T*-range", the structural relaxation time increases from 35 to 1300 in dimensionless units. We show the resulting PE distributions in Figure 6.

From the results shown in Figure 6 we find that the PE distributions for the bulk and string particles differ when compared for the same *T*, as was previously found by Ref. 63 for a binary Lennard-Jones mixture. In addition, we find that the PE distribution for particles in the bulk clearly vary with *T*, while particles moving in strings show PE distributions that vary significantly less with *T*. We further note that the PE distributions for particles moving in strings exhibit higher average values of PE relative to the bulk values at the same *T*, similar to what Ref. 63 observed for immobile/mobile particles.

Finding that the PE distributions for particles moving in strings do not shift substantially with temperature relative to the bulk suggests that the local string environment is in some way unique, persistent, and different from the bulk. To further test this finding we analyze the distribution of interparticle distances for particles in the bulk and strings, and test the *T*-dependence of this measure.

In Figure 7 we show for nine temperatures the probability, P(r), of finding a particle a distance *r* from an arbitrary reference particle in the bulk (top panel) and in the strings (lower panel). This plot shows that P(r) clearly depends on *T* for the bulk liquid, while it is independent of *T* for the particles moving in strings.

One can also plot the radial distribution function, g(r), and obtain the same results as with P(r), since P(r) is equal to g(r) less the normalizing to remove neighbors occurring in a completely random (ideal gas) system. In addition, g(r) contains information about the local structure that can be easily interpreted. For example, a split second peak is indicative of five-fold symmetry, which in turn is an indication of an underlying icosahedral structure.[64] Hence, we use g(r) to test if the local string environment resembles that of a high *T* bulk environment at any *T*. To do so we overlay high *T* bulk g(r) curves on the string ones for various high *T*, as shown in Figure 8 for $T = 0.65$, which we find to be the best fit between a high $T$ g(r) and the string g(r).

As shown in Figure 8, we find that the $T = 0.65$ g(r) bulk liquid curve merges with the string curves at $r \approx 2.7$ and that there is a good fit of the first peak. Interestingly, we find the first minimum for the bulk liquid at $T = 0.65$ to be deeper than that of the strings, while the second peak lacks the clear split that is observed for the strings. From these tests we find that there exist a unique local string environment that persist up to $r \approx 2.7$ away from a given string particle, and that the string environment is not that of the bulk liquid at any $T$. Rather, we deduce that the local string environment is a unique short-range environment that promotes string-like motion.

To further analyze the local string environment in the low $T$-range, we calculate the coordination numbers for particles in the bulk and in the strings by integrating the area under the first neighbor peaks in g(r). At $T = 0.42$ we find an average coordination number of 12.72 ± 0.05 for the bulk and 12.04 ± 0.05 for the strings. At $T = 0.52$ the average coordination number for the bulk is 12.52 ± 0.05, and 12.09 ± 0.05 for the strings. Hence, we find that particles in the bulk tend to have a higher coordination number relative to the string particles, and that this trend becomes more pronounced with decreasing $T$ for the bulk. In contrast, the average coordination number for strings is independent of $T$. The differences in coordination number can be interpreted as variations in local density: particles moving in strings have a lower local density relative to the bulk average density. Observations of density fluctuations have been made experimentally by Cui and co workers,[35] where such fluctuations were found to be the key stage for a particle to escape the cage of surrounding particles; the fluctuations were found to "fluidize"[65] the local environment of a particle that then was able to move on the order of one particle diameter by rapid diffusion. This scenario is qualitatively similar to what we find here in the DZ model for particles moving in strings.

Upon further investigation of the local structural environment of the string particles, we find that as the string particles move, the string-environment can be described as dominated by elongated cages. These cages can be described as constituted primarily of "stacked" rings, each comprised of five particles – hence the significant icosahedral nature of the string environment indicated by the split second peak in g(r) for the strings. We show an example of this environment in Figure 9, where a string particle is depicted with the surrounding *NNs* at two different time instances. The string particle is dark red at the initial time and dark blue at the final time. The *NN* particles are red at the initial time and light blue at the final time.

From Figure 9 we clearly see that the string particle is able to move a distance (about 1.5 particle diameters) that is significantly larger than the displacement for any of the *NN*. The particle, hence, moves in an elongated cage that is made up by *NN* that are more or less immobile over the time interval that the string particle moves. We find this scenario for particles moving in strings in the low *T*-range; that is, particles "jump" in elongated cages at low *T* in the DZ model. This finding is in agreement with that of Takeuchi obtained for a glassy polymer,[66] where the same scenario was observed for "jumping" monomers. Recently, in an exhaustive study, Jain and de Pablo[38] report a connection between local structure and particle dynamics in that particles with a high degree of distortion of the local environment (measured in terms of the Delaunay simplices[39]) are likely to transition faster between metabasins on the potential energy landscape. Similarly, in a preceding study, Starr et al.[40] found broad distributions of asphericity values for the Voronoi[39] volumes in a polymer melt, a result which is consistent with ours and with that of Ref. 38 with regards to the distortion of the local structural environment. We also note a recent study where hard spheres (HS) are used to model the dynamics of hard chains in a matrix of fixed HS,[67] in which the authors speculate that the distribution of void space and the connectivity of the void space may be important factors for the dynamics; these speculations agree well with our findings.

**VI. Common neighbor analysis**

To gain further insight as to the details of the local environment for the strings relative to the bulk liquid, we analyze the local structure in more detail than what is possible using the radial distribution function. Here we make use of the common neighbor (CN) analysis method.[6,43-45] CN is a measure of local structure, where a pair of particles is used as a "root pair" and their immediate surrounding local structure is characterized using a three-letter index. In this study we use a form of CN analysis developed by Jónsson and co-workers[44,45] where the three CN indices, $j$, $k$, $l$, have the following definitions: $j$ is the number of shared nearest neighbors for the root pair; $k$ is the number of bonds between the shared nearest neighbors, and $l$ is the number of bonds in the longest chain formed by the shared nearest neighbors for the root pair. Neighbors to the root pairs in this work refer to particles that reside within a distance less than that of the first minima of g(r) for each of the root pair particles. An example of CN indexing is given in Figure 10.

The CN method is simple but powerful enough to clearly distinguish between various local structures. For example, the only bonded pairs in the fcc crystal are represented by the index 421, while hexagonal close packing (hcp) has indices 421 and 422, and similarly, the perfect bcc crystal has only indices 666 and 444 present. Further, indices 555 signify pentagonal bipyramids, and a particle with 12 nearest neighbors exhibiting exclusively indices 555 is located at the center of an icosahedron. In addition, if a bond is broken between the 12 outer particles in an icosahedron, two of the 555 indices are transformed into 544 indices and two into 433 indices.[44] We can therefore use indices 544 and 433 as an indication of distorted local icosahedral order. Specifically, the 433 index arises from polytetrahedral structures consisting of three tetrahedra, while the 544 index denotes a structure consisting of four tetrahedra.[10]

Figure 11 shows CN results for the DZ system, for the bulk liquid in the top panel and for the strings in the bottom panel. We include data for $T = 1.6$ for the bulk in order to observe the distribution of CN indices in a very high $T$ liquid. This allows us to better determine the nature of the change in the distributions in the system as $T$ varies.

From the top panel in Figure 11 we clearly see that for the bulk the 555 index increases substantially in occurrence with decreasing $T$, indicating that the local structure becomes increasingly icosahedral in geometry. We also find that the other CN indices vary in occurrence over the $T$-range; the 433 and 544 indices show a mild maximum in values at $T \approx 0.6$, while the rest of the indices show a decrease in occurrence with decreasing $T$. The trend observed here for the index 555 is in agreement with previous studies of this model,[48] and shows that icosahedral SRO strongly increases with the degree of supercooling for this model system. In addition, the trends found here for the 555, 433, 422, and 421 indices closely match those found for a Lennard-Jones system.[6] When comparing our CN results for the bulk liquid with simulations where tight-binding and embedded-atom potentials are used to simulate lead[9] and aluminum,[10] respectively, one finds that those studies show slightly increasing values for the 544 and 433 indices with decreasing $T$. However, though the DZ and LJ[6] models show a slight decreasing trend for these indices with decreasing $T$, the occurrence percentage is in remarkably good agreement with the more element-specific studies. We note that Jakse and coworkers[68,69] have used *ab initio* MD to show that the SRO is more complex than just icosahedral in chemically detailed models. Still, despite the simplicity of the DZ and LJ systems, they are capable of representing the local structure of a metallic liquid to a high degree. We also note that the CN results for the DZ system agree well with the results of the particle type analysis carried out in this work, verifying that the various particle types are useful measures of local structure.

**VII. The low CN criterion**

When calculating the CN distributions for many string particles over consecutive time steps we find that particles moving in strings always exhibit unusual CN distributions at the onset of mobility. The observed CN distributions are always comprised of several indices reflective of very low local density. (By low local density we refer here specifically to CN indices that are "low", including 311 and lower, i.e. 300, 211, 200, 100, 000). Based on this observation, we now attempt to form a criterion, based solely on CN indices, that is indicative of an increased propensity for mobility, as compared to the average particle. This criterion is based on the three-letter CN index, *ijk*, and is as follows:

1) a particle must have three CN indices where $i \leq 3$, and

2) out of the three indices, two must have $j \leq 1$ and $k \leq 1$.

We will be referring to this criterion simply as the "LCN" criterion. For reference, we show in Figure 12 examples of configurations that exhibit low CN indices. An example of a particle with the LCN criterion fulfilled is shown in Figure 13.

The LCN particle shown in Figure 13 is indicated in red, and the upper leftmost panel shows the LCN enclosed by a cage of *NN* particles. The colors show the local structural environment of the *NN*, where dark blue indicates a particle belonging to a clustered icosahedra, pale blue a particle belonging to a single icosahedron, and white liquid-like particles. Upon rotating the image ~90° (we use Rasmol, www.openrasmol.org, for 3-D inspection of images), we find that the cage surrounding the LCN particle has a big hole where the red LCN particle completely lacks *NN* on one side, as shown in the upper rightmost panel. Upon an additional ~90° rotation (lower panel), we see that the LCN particle has no *NN* present in the cage from that side; the cage is open or "broken". This means that potentially the LCN particle could move in this direction. In addition, note that this space is also available for *NN* to the LCN particle if the *NN* resides such that it is bordering that space. This means that not only the LCN particle itself, but also a particle neighboring to the LCN particle, can potentially move into the available space.

Moreover, we note that when a string particle has become mobile and is translating through the liquid, it is not necessarily satisfying the LCN criterion at that time. This is due to the low coordination often found for already mobile string particles, as shown above. With a low enough coordination, space is available for translational motion and the LCN criterion of having three low CN indices may not be met simply due to the string particle having too few *NN*. Hence, we stress that LCN is related to the broken (open) cages needed for onset of mobility, not mobility in itself. In Figure 14 we illustrate this scenario by showing a region in space where a string will form. In the upper panel we show all particles in this region of space that satisfy the LCN criterion at some point over the time interval, $\Delta t^*$, over which strings are known to be of maximum length.[25] In the lower panel we show the same region with the string particles in red, superimposed on the LCN particles.

We can see from Figure 14 how the string extends through the region of LCN particles. In this example, all string particles were LCN particles at the onset of motion, with the exception of one particle that had a *NN* that was a LCN at the time for onset of motion.

To further quantify the local environment of the LCN particles with respect to structure, we compute a breakdown of the LCN particles in terms of the particle "types" defined above. We find that cage breaking, which LCN is indicative of, occurs mainly in the liquid-like region bordering the surfaces of the clusters of icosahedrally-ordered particles. This is indicated by the LCN particles being of particle type *F* in 68.7 ± 1.7 % of the cases. In comparison, the bulk-averaged value for type *F* is 41.5 ± 1.66 %, so the type *F* particles are clearly over-represented in the LCN distribution relative to the CN distribution of the bulk.

The complete distribution of particle types for the LCN particles at $T = 0.42$ is shown in Table I.

From Table I we also find that particles in icosahedral configurations are under-represented in the LCN distribution relative to the bulk, these are types *A* through *E*, while the particles deep in the liquid-like regions, type *G*, are represented with about the same frequency (within the error) as in the bulk averaged case.

**VIII. Properties of LCN particles in different model liquids**

In order to address the universality of the LCN criterion, we test for the existence of LCN particles in other glass-forming model liquids. Here we make use of the polymer model used by Bennemann and coworkers.[70] We choose this model since it has been successfully used in previous studies relating to SHD and SLM, hence, relevant information relating to our study of the origins of SHD/SLM is available for this model in the literature (see e.g. Ref. 32 and references therein). Further, as this polymer model is a one-component system, and the only difference relative to the DZ system is the particle connectivity in the polymer chains, we can analyze the polymer system in a manner identical to the DZ model. In addition to the polymer model, we also make use of the 80:20 binary Lennard-Jones (LJ) model introduced by Kob and Andersen.[71] This model is chosen as it is one of the most well known model systems for glass-forming liquids, and because such a wealth of general information regarding the properties of this model is already published. Moreover, as Donati and coworkers[25] used this model to introduce the notion of SLM, specific information pertinent specifically to SLM is also available in the literature.

We begin by establishing the percentage of particles that are LCN particles as a function of $T$ in the different models, and show the results in Figure 15. Note that as we only generated two data points for the LJ system to test for the existence of LCN particles, we refrain here from discussing the trend of the temperature dependence on the percentage of the LCN particles in the LJ system. In order to compare the models, we use a normalized temperature scale, $T/T_{MCT}$, where $T_{MCT}$ is the critical temperature of the mode coupling theory (MCT)[54] for each model. We use $T_{MCT} = 0.4$ for the DZ system,[49] $T_{MCT} = 0.32$ for the NVT polymer system,[70] and $T_{MCT} = 0.435$ for the LJ system.[71] All data is generated as described by the cited authors and the systems are equilibrated for a minimum of $10\tau_\alpha$ at each $T$. We use 200 chains each consisting of 10 beads, i.e. $N = 2000$, for the polymer, and $N = 8000$ for the LJ system.

From Figure 15 we find monotonic increases in the percentage of LCN particles with increasing $T$ for the DZ and polymer models. The increase in the number of LCN particles is a reflection of the decrease of the impact of caging on the dynamics with increasing $T$. This phenomenon is seen in the mean squared displacement (MSD) plots for supercooled liquids (see e.g. Ref 27 for a MSD plot for the DZ system), where the plateau that is characteristic of caging[71] vanishes with increasing $T$. The trend seen in the MSD plots for supercooled liquids reflects that, with increasing $T$, a larger number of particles reside within cages that are "broken" at any given time, a trend that is also detected using the LCN criterion. Hence, the results shown in Figure 15 reinforce the idea of using the LCN criterion as a measure to detect particles with a high propensity for mobility.

We also find from Figure 15 that the occurrence of LCN particles is larger for the LJ system relative to the other two models by a factor of about three, and that the DZ and the polymer models show similar values. This is very likely an effect of challenges involved in obtaining correct values for the three different cutoffs that are necessary to establish *NN* relationships in the binary liquid. It may also be that the LCN criterion must be obtained uniquely for the LJ model; the idea behind LCN is of course transferable between systems of different composition, but the rules may need to be modified between monatomic models and models with different particle sizes. We will address this issue later in this manuscript.

We used the polymer system to test for a connection between particles satisfying the LCN criterion and mobility in a system different from the monatomic DZ model. The result is identical to what we find for the DZ model; particles moving in strings in the polymer system either are LCN at the onset of mobility, or have *NN* that are LCN at that time. In Figure 16 we show an example from the polymer system, where images that demonstrate the connection between the LCN particles and the strings are displayed. Note that this figure is equivalent to that of the DZ model shown in Figure 14.

We next test for LCN clusters by connecting LCN particles that are *NN* to one another. The resulting probability distributions, *P(S)*, of the cluster sizes, *S*, are shown in Figure 17 together with fits that are of the following form: $P(S) = k_1 S^{-\varsigma} \exp[-(S/S_0)^{k_2}]$. In the fitting function, which is an expression comprised of a power law with an exponential cutoff, $\varsigma$ is the power law exponent, $S_0$ is a characteristic cluster size, and $k_1$ and $k_2$ are fitting parameters.

From Figure 17 we find that the LCN particles do form clusters in all model liquids, and that the distributions, with the exception of the high $T$ polymer data, fit well to Equation 6.1. We suspect that the shape of the distribution in the case of the high $T$ polymer data is due to finite size effects. Previous authors[32] have reported finite size effects for a smaller version ($N = 1200$) of this system. We use $N = 2000$, but finite size effects still seem to occur as the LCN clusters grow in size with increasing $T$. To continue the discussion of Figure 17, we summarize in Table II the parameter values obtained from the fits.

From Table II, we find that the power law exponent, $\varsigma$, which is reflective of the slope of the fits at smaller cluster sizes before the exponential cutoff begins to dominate the functional form of the fitting expression, is of similar magnitude in all cases. While fitting, we also noticed that $\varsigma$ is relatively insensitive to the values used for $S_0$, the characteristic cluster size, allowing us to get good estimates of the error in $\varsigma$. Hence, despite difference in the percentage of particles satisfying the LCN criterion in the LJ model relative to the other two models, as shown in Figure 15, no noticeable impact is found on the clustering of the LCN particles.

Further examining Figure 17, we find the LCN distributions for the polymer and the LJ systems reflect increasing cluster sizes with decreasing $T$. This is seen as the higher $T$ data in these cases begin to drop faster than the low $T$ data. Even though finite size effects affect the high $T$ polymer data, this trend is clearly detectable for $S < 11$ before the data becomes distorted. We note that that clusters of mobile particles grow in size with decreasing $T$ for these model systems,[27,63,72] and speculate that such a trend would be observed for the LCN clusters as well if the LCN criterion is related to mobility.

However, the observed trend for the LCN cluster size distributions for the DZ model is the opposite of what we expect if the LCN clusters mapped onto clusters of highly mobile particles, which grow in size with decreasing T. Here the high $T$ data indicate that larger LCN clusters form at higher $T$ rather than at low $T$. This observation is rationalized by the fact that the DZ model liquid is highly ordered in the low $T$-range (for $T \leq 0.52$). In this very cold liquid, clusters of icosahedrally-ordered particles (CIP) comprise up to ~50% of the particles at any given time. These CIP, hence, impose constraints on the liquid, and as shown above using the particle types, the LCN particles are found to reside mainly in the interface between the CIP and liquid-like regions. We show in Figure 18 an image from the simulation box at low $T$ for the DZ model depicting the described confinement. As shown in Figure 18, the LCN particles appear mainly in the interface between the CIP and the liquid-like regions. Hence, at low $T$ the LCN particles cannot form large clusters, simply due to the physical constraints imposed by the CIP. At higher $T$, the impact of the CIP decreases, and the LCN particles can form clusters that are larger in size. We note that a decrease in size of clusters of high potential-energy propensity particles with decreasing $T$ was recently reported by Matharoo et al.[41] in ST2 water. As water is a highly structured liquid at low $T$, similar constraints may be occurring there.

From Figure 18, the LCN particles (green) can be observed to reside most frequently in the proximity of the clusters of icosahedrally-ordered particles (blue), in agreement with data presented in Table I above. Hence, as mobile clusters are defined to exist over a time interval $\Delta t$, which is typically taken as $\Delta t^*$ near the cage-breaking time, rather than at one snapshot in time as is the case of LCN particles, all particles in mobile clusters can be LCN particles at some point in the given time interval. However, as the mobile particles are not LCN at the same time, the LCN clusters in the DZ model become smaller than the mobile ones as $T$ decreases.

The LCN distribution observed for the DZ model in the low $T$-range is now fitted to an exponential as we note that the form observed for the distributions of string-lengths is exponential in all three model liquids; for the DZ,[27] the polymer,[32] and also for the LJ[25] model. We show the LCN cluster size distribution for the low $T$-range DZ model in Figure 19 plotted on a semi-log scale, where a straight line corresponds to an exponential.

Clearly, the full range of the data shown in Figure 19 do not fit to a single exponential. However, for values of $S < 8$ the data are well fit to an exponential $P(S) = A \exp[-S/S_0]$, with $A = 0.63$ and $S_0 = 2.6 \pm 0.17$. In Figure 2 we show a fit of the string lengths to an exponential with $L_0 = 2.5 \pm 0.18$. The fact that $S_0$ (LCN cluster size) and $L_0$ (string length) are in such good agreement is suggesting a close connection between the LCN particles and the micro strings, which is a subset of strings that move coherently in a single jump.[27] Specifically, as Ref. 27 showed that the coherently moving micro strings most often consist of 7 particles or less, it seems reasonable that the small LCN clusters are intimately related to SLM. Intuitively this makes sense, since LCN particles are particles with broken cages at a given time, and micro strings are particles moving together and breaking out of their cages simultaneously.[27]

**IX. Summary and conclusion**

To summarize, from MD simulations of the supercooled Dzugutov system, we found that the local structure in the vicinity of particles moving in strings is different at low $T$ from any structure found for the bulk liquid, and that string-like particle motion most frequently occurs in the boundary between clusters of icosahedrally-ordered particles and disordered, liquid-like domains. Clusters of icosahedrally-ordered particles are found to appear in significant amounts at the onset $T$ for non-Arrhenius dynamics. We also found that the environment of string particles varies minimally with $T$ in a low $T$-range, unlike that of the bulk liquid where a clear $T$-dependence on several structural quantities is observed. Further, in the low $T$-range, string particles are found to exhibit lower coordination numbers more frequently, and to have less icosahedral SRO relative to particles in the bulk. We also found that the local environment of particles moving in strings at low $T$ consists of elongated $NN$ cages.

Making use of the CN analysis method[44,45] we proposed a criterion that detects particles with open or "broken" $NN$ cages. These particles with open cages are termed LCN particles, as they all exhibit several unusually "low" CN indices. By investigating the DZ,[42] a polymer,[70] and a LJ[71] model, we showed that LCN particles are, in general, more frequently occurring as $T$ increases on heating from $T_{MCT}$ for all three model liquids, suggesting that all three liquids contain low-mobility clusters of locally preferred packing (icosahedral in the case of Dzugutov) coexisting with fluid-like regions.[60,61] The existence of LCN particles in all three liquids is reflective of the decreased impact of caging[71] on the dynamics with increasing $T$, a trend that is commonly observed for supercooled liquids. Further, we found that the LCN particles form clusters that can be fitted to a function comprised of a power law with an exponential cutoff. From fitting the distributions we found that the power law exponent is similar between the models in the limited $T$-range investigated here. The LCN distributions are very similar in form to the size distributions for mobile particle clusters published in the literature for all three model liquids,[27,63,72] and the power law exponents are of similar magnitude. For low $T$, i.e. $0.42 \leq T \leq 0.52$ in the case of the DZ model, we found that the LCN particles reside mainly in liquid-like regions interfacing clusters of icosahedrally-ordered particles – a result arising from frustration induced in those regions by the CIP.[62] Further, we showed that the LCN cluster size distribution for the DZ model at $T = 0.42$ exhibits exponential decay that for small cluster sizes, $S < 8$, closely matches that of the distribution of string lengths found at $T = 0.42$; an indication of a close connection between mobile particles moving coherently in micro strings and LCN particles.

We conclude from the similarities in properties found for particles satisfying the LCN criterion in the three model liquids that the LCN criterion can be universally implemented in particle-based model liquids to identify particles with a propensity for high mobility, and we do establish such a connection for the DZ and the polymer systems.

Our picture of the origin of string-like motion in the Dzugutov liquids is thus as follows. As $T$ decreases upon cooling, clusters of icosahedrally-ordered, relatively immobile particles appear in the liquid. In the interfacial region between the slow, ordered domains and disordered, liquid-like, regions, particle environments become geometrically frustrated, resulting in an elongation of particle cages due to the surrounding clusters and a concomitant low number of neighbors. This structural environment sets the stage for subsequent particle motion, wherein particles follow the direction of the long axis of the elongated cages. Because these cages are clustered together, multiple particles may follow one another in strings. As the liquid is further cooled and the low-mobility, icosahedral clusters grow in size, the interfacial region becomes more prevalent and longer lived, resulting in longer and more abundant strings. The low CN picture appears to hold in the polymer melt and LJ mixture as well, but with a potentially different local structural motif. This scenario is consistent with the facilitation model for glass formers[73,74] which we tested in the Dzugutov system in a previous publication,[50] as well as the frustration-limited domain model[60,61] and the fluidized domain model of spatially heterogeneous dynamics.[65]

**Acknowledgments**

We thank R. Ziff and A. Keys for stimulating discussions, and also A. Keys for assisting with figures 3 and 5. We gratefully acknowledge funding from NASA and the Dept. of Education GAANN Fellowship Program.

**References**

[1]H. Sillescu, J. Non-Crys. Solids, **243**, 81-108, (1999).

[2]M. D. Ediger, Annu. Rev. Phys. Chem., **51**, 99-128, (2000).

[3]P. G. Debenedetti and F. H. Stillinger, Nature, **410**, 259-267, (2001).

[4]H. C. Andersen, Proc. Natl. Acad. Sci. U. S. A., **102**, 6686-6691, (2005).

[5] F. C. Frank, Proc. R. Soc. London A, **215**, 43-46, (1952).

[6] H. Jónsson and H. C. Andersen, Phys. Rev. Lett., **60**, 2295-2298, (1988).

[7] P. J. Steinhardt; D. R. Nelson and M. Ronchetti, Phys. Rev. Lett., **47**, 1297-1300, (1981).

[8] P. J. Steinhardt; D. R. Nelson and M. Ronchetti, Phys. Rev. B, **28**, 784-805, (1983).

[9] L. Hui and F. Pederiva, Chem. Phys., **304**, 261-271, (2004).

[10] M. Kreth; P. Entel; K. Kadau and R. Meyer, Phase Transit., **77**, 89-100, (2004).

[11] H. Reichert; O. Klein; H. Dosch; M. Denk; V. Honklmaki; T. Lippmann and G. Reiter, Nature, **408**, 839-841, (2000).

[12] T. Schenk; D. Holland-Moritz; V. Simonet; R. Bellissent and D. M. Herlach, Phys. Rev. Lett., **89**, (2002).

[13] T. Schenk; V. Simonet; D. Holland-Moritz; R. Bellissent; T. Hansen; P. Convert and D. M. Herlach, Europhys. Lett., **65**, 34-40, (2004).

[14] D. Holland-Moritz; T. Schenk; P. Convert; T. Hansen and D. M. Herlach, Meas. Sci. Technol., **16**, 372-380, (2005).

[15] M. M. Hurley and P. Harrowell, Phys. Rev. E, **52**, 1694-1698, (1995).

[16] A. Heuer; M. Wilhelm; H. Zimmermann and H. W. Spiess, Phys. Rev. Lett., **75**, 2851-2854, (1995).

[17] M. T. Cicerone and M. D. Ediger, J. Chem. Phys., **103**, 5684-5692, (1995).

[18] A. Heuer and K. Okun, J. Chem. Phys., **106**, 6176-6186, (1997).

[19] W. Kob; C. Donati; S. J. Plimpton; P. H. Poole and S. C. Glotzer, Phys. Rev. Lett., **79**, 2827-2830, (1997).

[20] C. Donati, S.C. Glotzer, and P.H. Poole, Phys. Rev. Lett. **82**, 5062, (1999); C. Bennemann; C. Donati; J. Baschnagel and S. C. Glotzer, Nature, **399**, 246-249, (1999).

[21] R. Richert, J. Phys.: Cond. Matter, **14**, R703-R738, (2002).

[22] J. Baschnagel and F. Varnik, J. Phys.-Condes. Matter, **17**, R851-R953, (2005).

[23] E. R. Weeks; J. C. Crocker; A. C. Levitt; A. Schofield and D. A. Weitz, Science, **287**, 627-631, (2000).

[24] W. K. Kegel and A. van Blaaderen, Science, **287**, 290-293, (2000).

[25] C. Donati; J. F. Douglas; W. Kob; S. J. Plimpton; P. H. Poole and S. C. Glotzer, Phys. Rev. Lett., **80**, 2338-2341, (1998).

[26] T. B. Schroder; S. Sastry; J. C. Dyre and S. C. Glotzer, J. Chem. Phys., **112**, 9834-9840, (2000).

[27] Y. Gebremichael; M. Vogel and S. C. Glotzer, J. Chem. Phys., **120**, 4415-4427, (2004).

[28] R. Zangi and S. A. Rice, Phys. Rev. Lett., **92**, (2004).

[29] N. Giovambattista; F. W. Starr; F. Sciortino; S. V. Buldyrev and H. E. Stanley, Phys. Rev. E, **65**, 041502, (2002).

[30] N. Giovambattista; S. V. Buldyrev; H. E. Stanley and F. W. Starr, Phys. Rev. E, **72**, (2005).

[31] R. Yamamoto and A. Onuki, Phys. Rev. E, **58**, 3515-3529, (1998).

[32] M. Aichele; Y. Gebremichael; F. W. Starr; J. Baschnagel and S. C. Glotzer, J. Chem. Phys., **119**, 5290-5304, (2003).

[33] M. Vogel and S. C. Glotzer, Phys. Rev. Lett., **92**, 255901, (2004).

[34] A. H. Marcus; J. Schofield and S. A. Rice, Phys. Rev. E, **60**, 5725-5736, (1999).

[35] B. X. Cui; B. H. Lin and S. A. Rice, J. Chem. Phys., **114**, 9142-9155, (2001).

[36] E. R. Weeks and D. A. Weitz, Phys. Rev. Lett., **89**, (2002).

[37] A. Widmer-Cooper; P. Harrowell and H. Fynewever, Phys. Rev. Lett., **93**, (2004).

[38] T. S. Jain and J. J. de Pablo, J. Chem. Phys., **122**, (2005).

[39] M. Tanemura; T. Ogawa and N. Ogita, J. Comput. Phys., **51**, 191-207, (1983).

[40] F. W. Starr; S. Sastry; J. F. Douglas and S. C. Glotzer, Phys. Rev. Lett., **89**, (2002).

[41] G. Matharoo; M. Razul and P. Poole, cond-mat/0607014, (2006).

[42] M. Dzugutov, Phys. Rev. A, **46**, R2984-R2987, (1992).

[43] J. D. Honeycutt and H. C. Andersen, J. Phys. Chem., **91**, 4950-4963, (1987).

[44] A. S. Clarke and H. Jónsson, Phys. Rev. E, **47**, 3975-3984, (1993).

[45] D. Faken and H. and Jónsson, Comp. Mat. Sci., **2**, 279-286, (1994).

[46] M. Dzugutov, Phys. Rev. Lett., **70**, 2924-2927, (1993).

[47] J. P. K. Doye; D. J. Wales and S. I. Simdyankin, Faraday Discuss., **118**, 159-170, (2001).

[48] F. H. M. Zetterling; M. Dzugutov and S. I. Simdyankin, J. Non-Crys. Solids, **293**, 39-44, (2001).

[49] M. Dzugutov; S. I. Simdyankin and F. H. M. Zetterling, Phys. Rev. Lett., **89**, (2002).

[50] M. N. J. Bergroth; M. Vogel and S. C. Glotzer, J. Phys. Chem. B, **109**, 6748-6753, (2005).

[51] J. Roth and A. R. Denton, Phys. Rev. E, **61**, 6845-6857, (2000).

[52] H. J. C. Berendsen; J. P. M. Postma; W. F. Vangunsteren; A. Dinola and J. R. Haak, J. Chem. Phys., **81**, 3684-3690, (1984).

[53] W. C. Swope; H. C. Andersen; P. H. Berens and K. R. Wilson, J. Chem. Phys., **76**, 637-649, (1982).

[54] W. Gotze and L. Sjogren, Rep. Prog. Phys., **55**, 241-376, (1992).

[55] J. Roth and F. Gahler, Eur. Phys. J. B, **6**, 425-445, (1998).

[56] Y. Gebremichael; M. Vogel; M. N. J. Bergroth; F. W. Starr and S. C. Glotzer, J. Phys. Chem. B, **109**, 15068-15079, (2005).

[57] G. Adam and J. H. Gibbs, J. Chem. Phys., **43**, 139-&, (1965).

[58] J. Dudowicz; K. F. Freed and J. F. Douglas, J. Chem. Phys., **111**, 7116-7130, (1999).

[59] A. Keys and S. C. Glotzer, Manuscript in preparation (2006).

[60] D. Kivelson; S. A. Kivelson; X. L. Zhao; Z. Nussinov and G. Tarjus, Physica A, **219**, 27-38, (1995).

[61] G. Tarjus; S. A. Kivelson; Z. Nussinov and P. Viot, J. Phys.-Condes. Matter, **17**, R1143-R1182, (2005).

[62] M. Hoare, Ann. NY Acad.Sci., **279**, 186-207, (1976).

[63] C. Donati; S. C. Glotzer; P. H. Poole; W. Kob and S. J. Plimpton, Phys. Rev. E, **60**, 3107-3119, (1999).

[64] D. R. Nelson, Phys. Rev. B, **28**, 5515-5535, (1983).

[65] F. H. Stillinger and J. A. Hodgdon, Phys. Rev. E, **50**, 2064-2068, (1994).

[66] H. Takeuchi, J. Chem. Phys., **93**, 2062-2067, (1990).

[67] R. W. Chang and A. Yethiraj, Phys. Rev. Lett., **96**, (2006).

[68] N. Jakse and A. Pasturel, Phys. Rev. Lett., **91**, (2003).

[69] N. Jakse; O. Le Bacq and A. Pasturel, J. Chem. Phys., **123**, (2005).

[70] C. Bennemann; W. Paul; K. Binder and B. Dunweg, Phys. Rev. E, **57**, 843-851, (1998).

[71] W. Kob and H. C. Andersen, Phys. Rev. E, **51**, 4626-4641, (1995).

[72] Y. Gebremichael; T. B. Schroder; F. W. Starr and S. C. Glotzer, Phys. Rev. E, **64**, (2001).

[73] J. P. Garrahan and D. Chandler, Proc. Natl. Acad. Sci. U. S. A., **100**, 9710-9714, (2003).

[74] J. P. Garrahan and D. Chandler, Phys. Rev. Lett., **89**, 035704, (2002).

**Tables**

| Particle type | A | B | C | D | E | F | G |
|---|---|---|---|---|---|---|---|
| **LCN (%)** | 0 | 5.8 | 0 | 18.1 | 0.4 | 68.7 | 6.4 |
| **STDEV (%)** | 0 | 0.7 | 0 | 1.5 | 0.2 | 1.7 | 1.3 |
| **Bulk averaged** | 0.8 | 8.9 | 6.0 | 35.0 | 3.7 | 41.5 | 4.2 |
| **STDEV (%)** | 0.1 | 0.9 | 0.6 | 1.8 | 1.0 | 1.7 | 0.8 |

**Table I**. The distribution of LCN particles in terms of types that reflect the local structural environment. Most LCN particles reside in the liquid-like region bordering domains of icosahedrally-ordered particles. See text for a more detailed discussion.

| Model | T/T$_{MCT}$ | $\varsigma$ | S$_0$ | $k_1$ | $k_2$ |
|---|---|---|---|---|---|
| DZ | 1.05 | 1.22 ± 0.09 | 9 | 1 | 1 |
| DZ | 1.37 | 1.43 ± 0.13 | 60 | 0.71 | 0.964 |
| Polymer | 1 | 1.67 ± 0.13 | 100 | 0.96 | 3 |
| Polymer | 1.09 | 1.86 ± 0.12 | 150 | 1.24 | 10 |
| LJ | 1.05 | 1.68 ± 0.14 | 180 | 1 | 1 |
| LJ | 1.08 | 1.83 ± 0.12 | 60 | 1.3 | 1.28 |

**Table II**. Parameter values obtained from fitting the LCN cluster size distributions, *P(S)*, of the three models to $P(S) = k_1 S^{-\varsigma} \exp[-(S/S_0)^{k_2}]$. See text for details.

**Figure captions**

Figure 1. The Dzugutov potential plotted together with the Lennard-Jones potential.

Figure 2. The distributions of string lengths at different *T* are shown together with exponential fits to the data. The fits are of the form P(S) ~ exp[-L/L$_0$]. L$_0$ for *T* = 0.42 is 2.5 ± 0.18, for *T* = 0.8 we find L$_0$ = 1.2 ± 0.08, and for *T* = 1.0 we find L$_0$ = 0.95 ± 0.06. The data indicate that longer strings appear with increased probability at lower *T*.

Figure 3. A typical example of a cluster of 98 icosahedrally-ordered particles in the equilibrium bulk liquid at *T* = 0.42.

Figure 4. The distribution of particle types in the bulk liquid (top panel) shows particles residing in icosahedrally-ordered clusters (*C*, *D*, and *E*) occurring more frequently as *T* is lowered, while the opposite trend is shown for liquid-like particles (*F* and *G*). Little dependence on *T* is found for *A* and *B*, indicating that single icosahedra are always present in the system. For the strings (lower panel), the distribution is dominated by *F* particles, indicating that strings occur most often in the vicinity of clusters of icosahedrally-ordered particles. This trend is enhanced with decreasing *T* as both *D* and *F* particles increase, while *G* decreases, in number at lower *T*.

Figure 5. A string (in red) extending through a liquid-like domain (light grey), located between clusters of icosahedrally-ordered particles (blue).

Figure 6. Distributions of potential energy (PE) for particles in the bulk (top panel) show a dependence on *T* that is strong relative to that of the strings (lower panel), which exhibit a mild dependence on *T*. Also shown is that the PE distributions for the strings are shifted to higher PE relative to the bulk, consistent with the findings of Ref. 63, which performed a similar analysis of highly mobile particles.

Figure 7. Probability, P(r), of finding an arbitrarily chosen particle a distance *r* from a given reference particle. This measure varies with *T* for the bulk liquid (top panel), while it is independent of *T* for strings (lower panel). Bulk data is included in the plot for $T = 0.42$, 0.43, 0.46, 0.49, 0.52, 0.55, 0.65, 0.80, and 1.0. String data is plotted at $T = 0.42$, 0.43, 0.46, 0.49, and 0.52.

Figure 8. By overlaying the radial distribution function, g(r), for the strings with g(r) for the bulk liquid at various *T*, we find that the local environment for particles moving in strings differs from that of the high *T* bulk liquid. The best fit is found for $T = 0.65$, although no temperature is a good match. See text for details.

Figure 9. A string particle, labeled S, is shown together with the surrounding *NN* at two different times. The string particle is dark red at the initial time and dark blue at the final time. The *NN* particles are red at the initial time and light blue at the final time. The scale shows the particles as smaller than actual size, the string particle moves about 1.5 particle diameters. $T = 0.42$.

Figure 10. Example of a structure that is represented by a CN index of 433. Here a root pair (grey particles) share 4 common neighbors (white particles), and the shared neighbors are connected via 3 "bonds" forming a chain of length 3.

Figure 11. The distribution of CN indices shows how the bulk liquid (top panel) becomes increasingly icosahedrally-ordered as *T* decreases, a result reflected by the rapid increase in 555 indices with decreasing *T*. The distribution of CN indices for the strings differs from that of the bulk over the 0.42 to 0.52 *T*-range as it shows minimal dependence on *T*. Also, for the strings, the 433 index is most frequently occurring, indicating a more disordered local environment relative to the significantly more ordered bulk in this low *T*-range.

Figure 12. Examples of structures that are identified using the CN analysis method to possess low indices. The root pair is indicated as grey particles, and the shared nearest neighbors are shown in white. All lines between particles indicate "bonds".

Figure 13. A particle (red) detected using the LCN criterion (see text for a definition) shown in three different views. The leftmost view shows the red LCN particle surrounded by all *NN* in a fashion that seem to enclose the particle in a cage. However, when rotating the image ~ 90° (middle panel), we see that the cage appears to be open on the right-hand side. Further, when rotating the image another 90° (rightmost panel) we find that the cage is wide open as the red LCN particle is visible without obstruction of any of the *NN*. The colors indicate the local structural environment of the *NN*, where dark blue indicates clustered icosahedra, pale blue a single icosahedron, and white liquid-like particles. The result is from the DZ model at $T = 0.42$.

Figure 14. A region in space over a time interval where a string develops is shown. The upper panel is the region with LCN particles (green) only, and the lower panel has the string superimposed in red. In this example, all but one of the string particles were LCN at the time they became mobile. The non-LCN string particle, which is the one closest to the bottom, has a LCN as a nearest neighbor at the onset of mobility. Data for $T = 0.42$.

Figure 15. The frequency of occurrence of LCN particles in the different model systems plotted as a function of $T/T_{MCT}$. Note that the LJ system exhibits significantly more LCN at the low *T* relative to the one-component DZ and polymer models. Monotonic increases in the number of LCN particles are observed for the DZ and the polymer model.

Figure 16. An example illustrating the connection between LCN particles and mobility in the polymer system. Shown in the upper panel are the LCN particles (green) that exist in a region of space where a string moves, at the times that the string particles become mobile. In the lower panel, the string particles (red) are superimposed on the LCN particles at the mobility onset times. Hence, we find for the polymer system, that particles moving in strings are either LCN themselves or have nearest neighbors that are LCN at the time for onset of mobility. Note that this figure is the equivalent of Figure 14 for the DZ model.

Figure 17. The distribution of LCN clusters for the DZ, polymer, and LJ models plotted on a log-log scale together with fits to the data. See text for a detailed discussion.

Figure 18. Shown here is a slice of the simulation box from the low *T* DZ liquid showing LCN particles in green, mobile particles in red, and particles in icosahedrally-ordered clusters in blue. The liquid-like particles are removed to ease visual inspection.

Figure 19. Plotted on a semi-log scale is the LCN cluster size distribution for the low *T*-range in the DZ model. The fit for a small initial region ($S < 8$) show good agreement to the fit of string lengths shown in Figure 2, indicating a close relationship between the LCN particles and micro strings. See text for details.

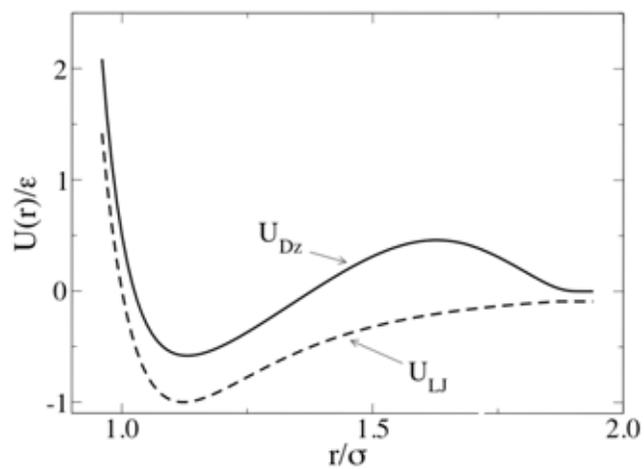

Figure 1.

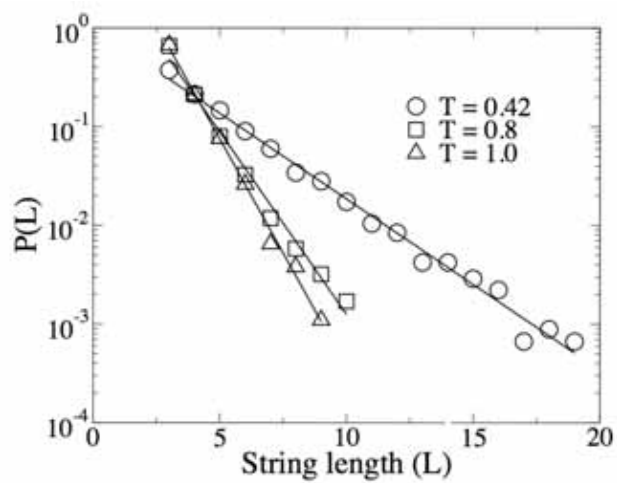

Figure 2.

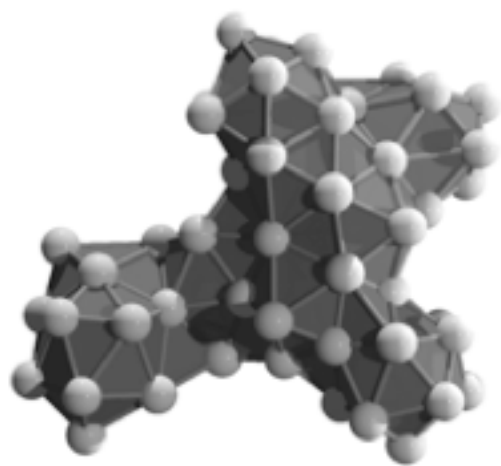

Figure 3.

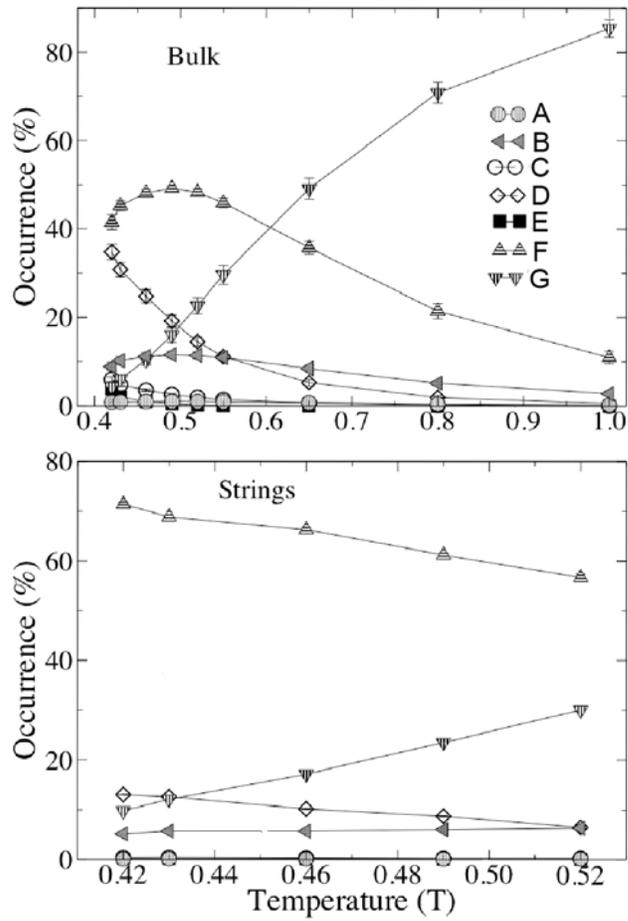

Figure 4.

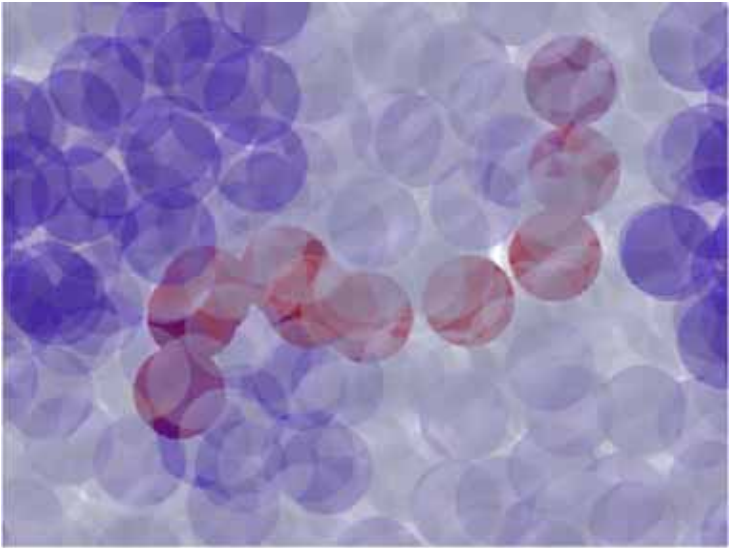

Figure 5.

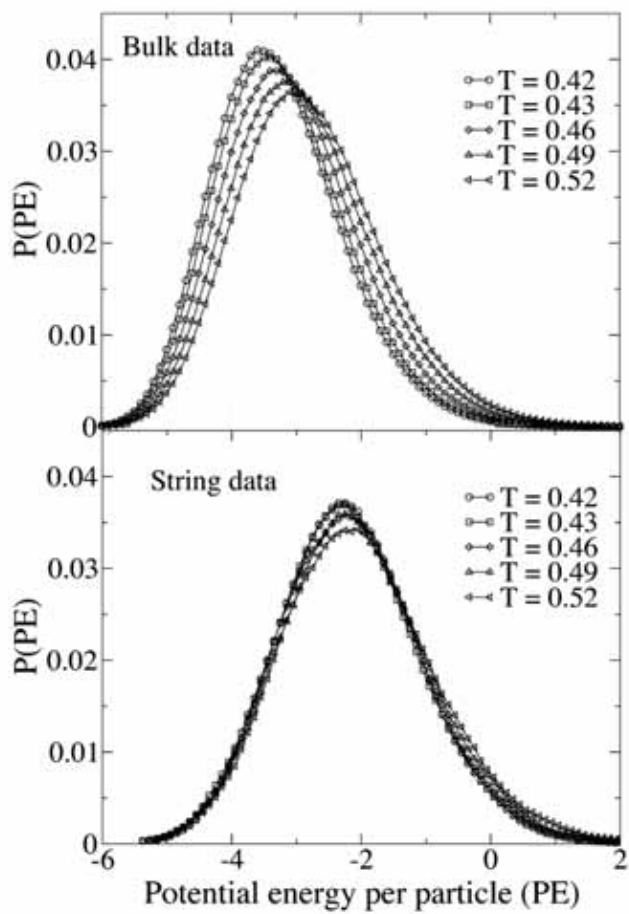

Figure 6.

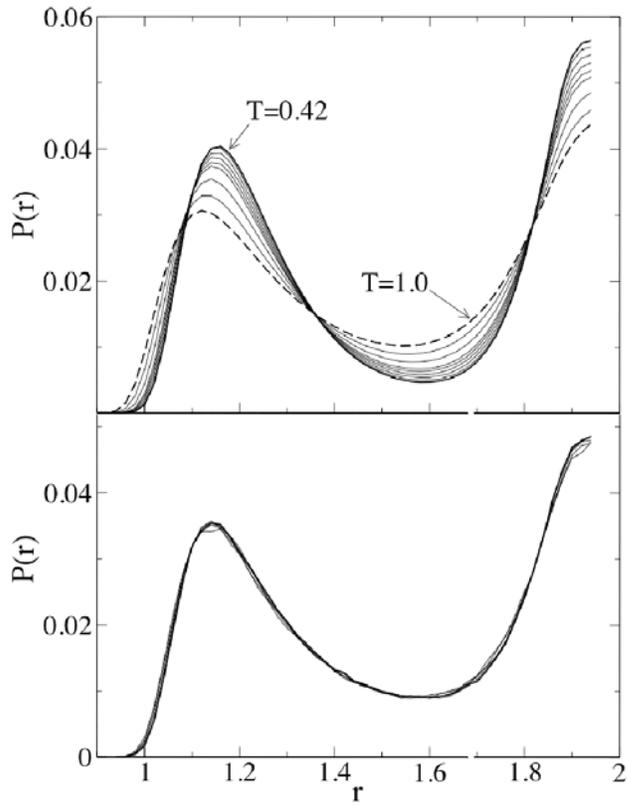

Figure 7.

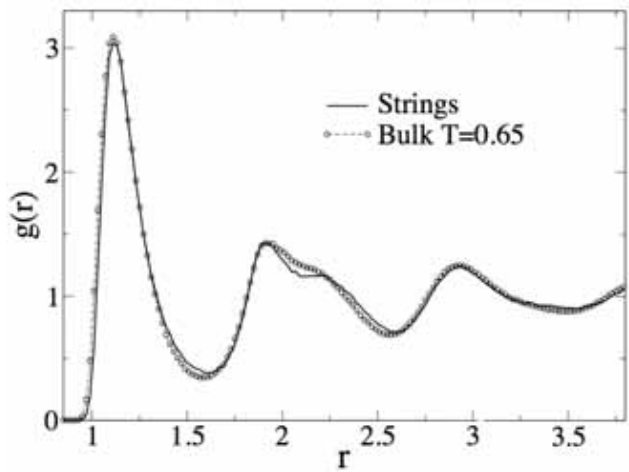

Figure 8.

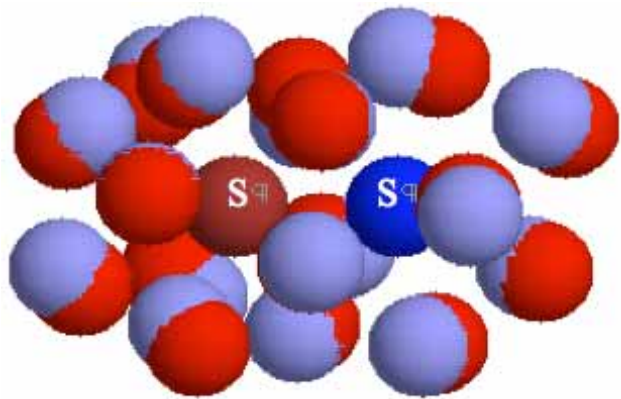

Figure 9.

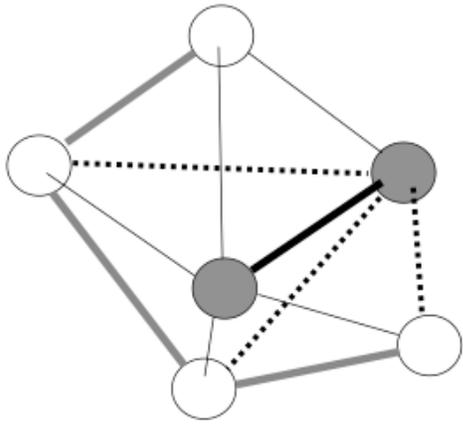

Figure 10.

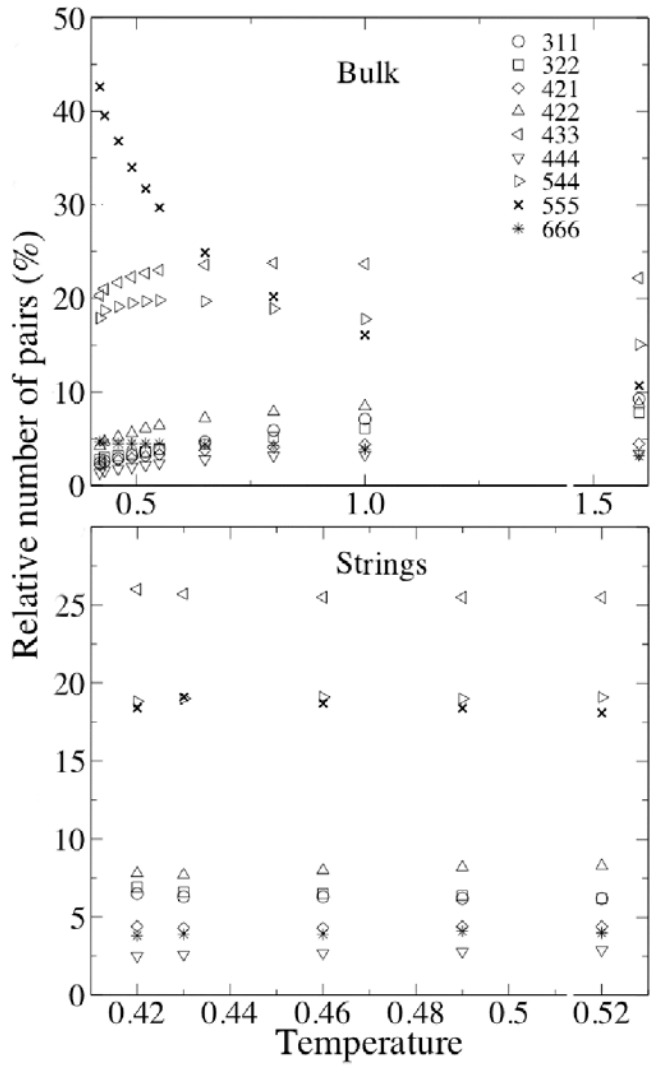

Figure 11.

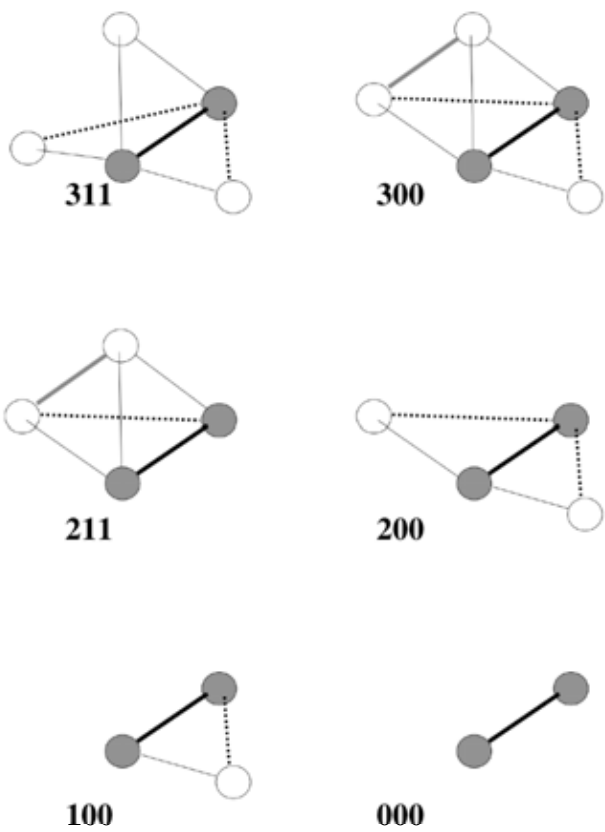

Figure 12.

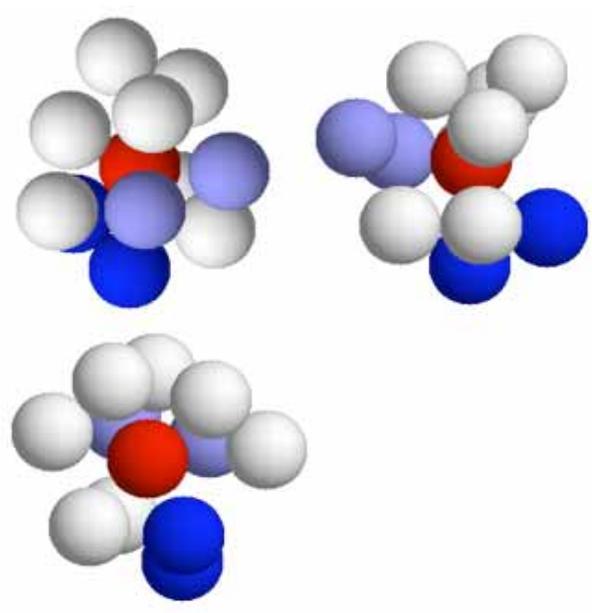

Figure 13.

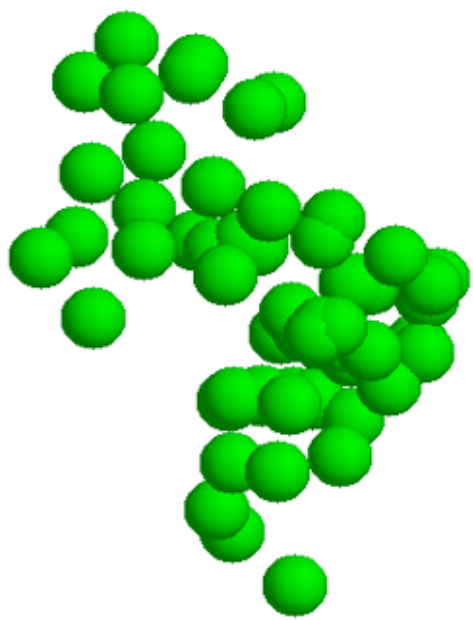

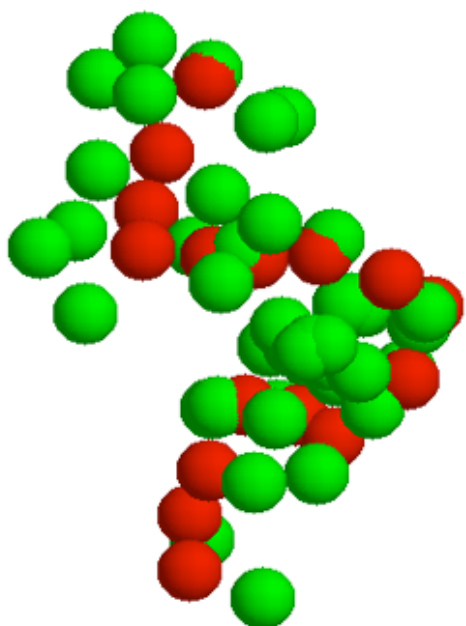

Figure 14.

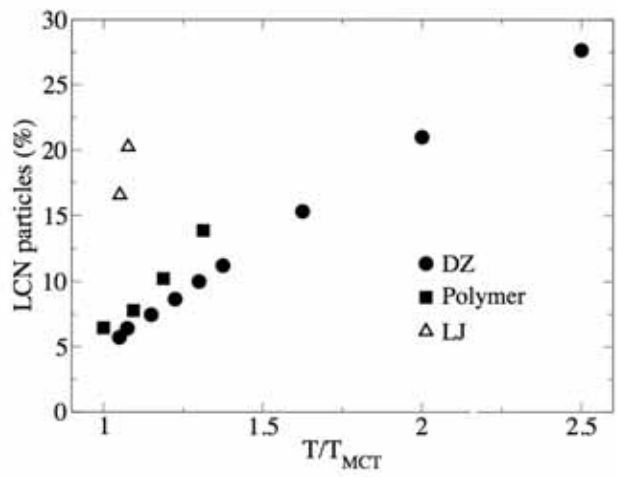

Figure 15.

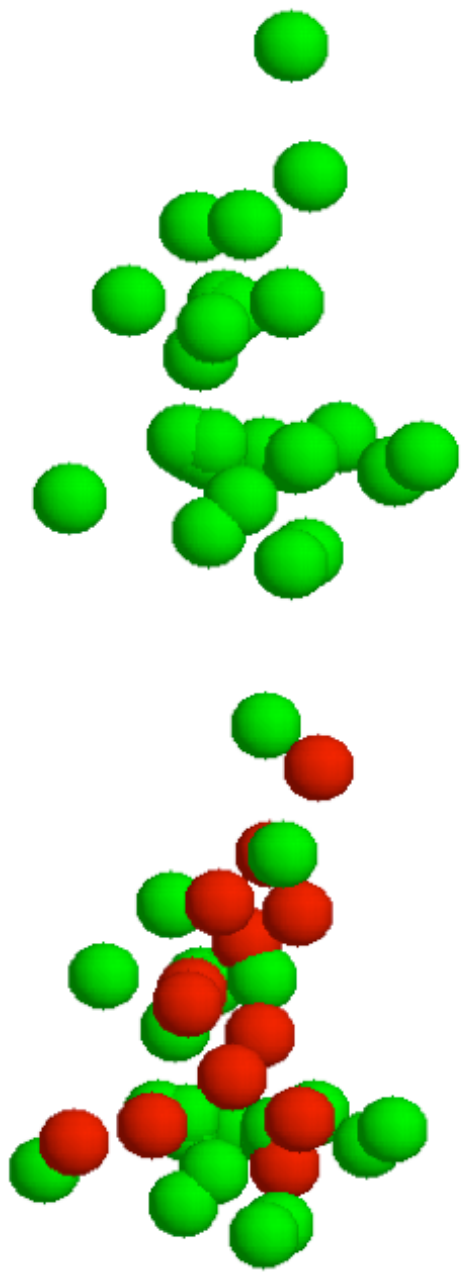

Figure 16.

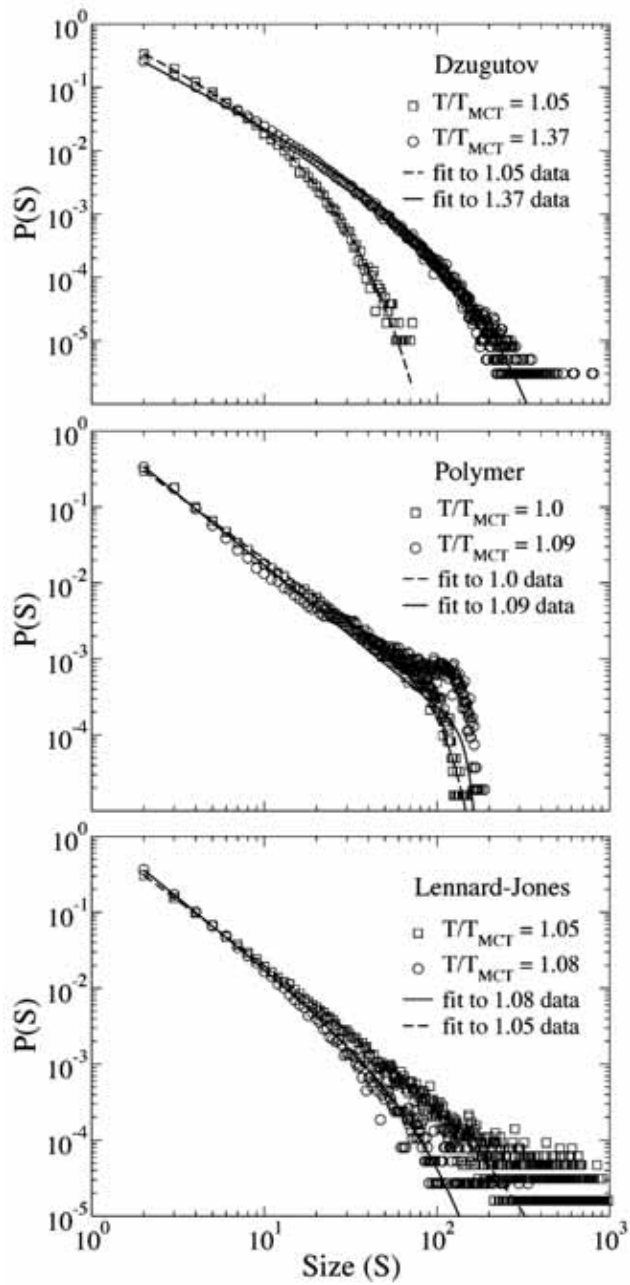

Figure 17.

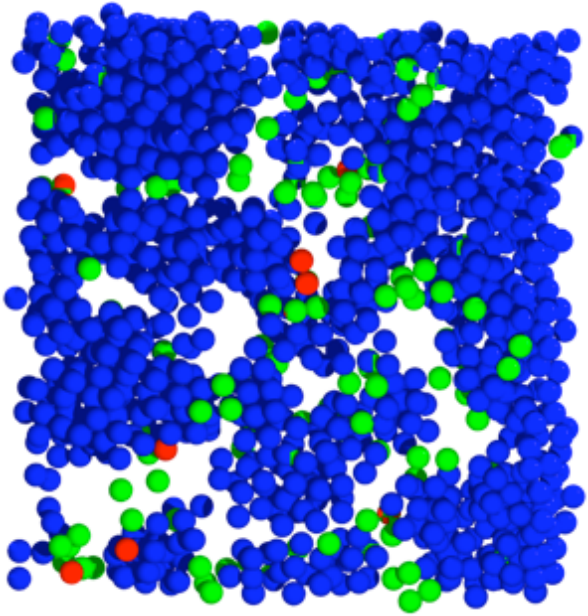

Figure 18.

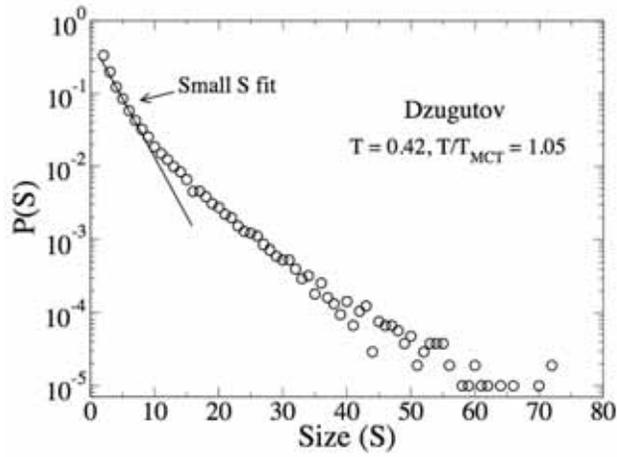

Figure 19.